\documentclass[sigconf]{acmart}

\usepackage{enumitem}
\usepackage{graphicx}
\usepackage{hyperref}

\AtBeginDocument{%
  }

\copyrightyear{2025}
\acmYear{2025}
\setcopyright{rightsretained}
\acmConference[MemSys '25]{International Symposium on Memory Systems}{October 07--08, 2025}{Washington, VA, USA}
\acmBooktitle{International Symposium on Memory Systems (MemSys '25), October 07--08, 2025, Washington, VA, USA}
\acmPrice{}
\acmDOI{10.1145/3767110.3767119}
\acmISBN{979-8-4007-2002-4/25/10}

\makeatletter
\renewcommand\footnotetextcopyrightpermission[1]{}
\makeatother

\begin{document}


\title{
  Secure IVSHMEM: End-to-End Shared-Memory Protocol with Hypervisor-CA Handshake and In-Kernel Access Control }


\author{Hyunwoo Kim}
\affiliation{%
  \institution{Intel}
  \city{Seoul}
  \country{South Korea}
}
\email{onion.kim@intel.com}

\author{Jaeseong Lee}
\affiliation{%
  \institution{Intel}
  \city{Seoul}
  \country{South Korea}
}
\email{jerry.j.lee@intel.com}

\author{Sunpyo Hong}
\affiliation{%
  \institution{Intel}
  \city{Seoul}
  \country{South Korea}
}
\email{brandon.hong@intel.com}

\author{Changmin Han}
\affiliation{%
  \institution{Intel}
  \city{Seoul}
  \country{South Korea}
}
\email{duke.han@intel.com}

\renewcommand{\shortauthors}{Kim et al.}

\begin{abstract}

  In-host shared memory (IVSHMEM) enables high-throughput, zero-copy
  communication between virtual machines, but today's implementations lack any
  security control, allowing any application to eavesdrop or tamper with the
  IVSHMEM region. This paper presents \textit{Secure IVSHMEM}, a protocol that
  provides end-to-end mutual authentication and fine-grained access enforcement
  with negligible performance overhead. We combine three techniques to ensure
  security: (1) channel separation and kernel module access control, (2)
  hypervisor-mediated handshake for end-to-end service authentication, and (3)
  application-level integration for abstraction and performance mitigation. In
  microbenchmarks, Secure IVSHMEM completes its one-time handshake in under
  100~$\mu$s and sustains data-plane round-trip latencies within 5\% of the
  unmodified baseline, with negligible bandwidth overhead. We believe this design
  is ideally suited for safety and latency-critical in-host domains, such as
  automotive systems, where both performance and security are paramount.

\end{abstract}

\maketitle




\section{Introduction}
The automotive industry is rapidly evolving, driven by advances in
semiconductor technology that have shifted system architectures from
traditional microcontrollers (MCUs) to powerful Systems-on-Chip (SoCs). This
evolution not only enhances computational capabilities but also paves the way
for Software-Defined Vehicles (SDVs), where flexibility, scalability, and rapid
updates are paramount. In SDVs, virtualization technology plays a crucial role
by enabling the coexistence of multiple virtual machines (VMs) on a single
hardware platform, ensuring isolated yet efficient execution of diverse
applications. For example, modern cockpit domain controllers often deploy
separate VMs for real-time operations (RTOS) and infotainment systems, which is
essential for balancing performance and
safety\cite{jiang2024towards,reinhardt2014embedded}.

\begin{figure}[ht]
  \begin{center}
    \includegraphics[width=\linewidth]{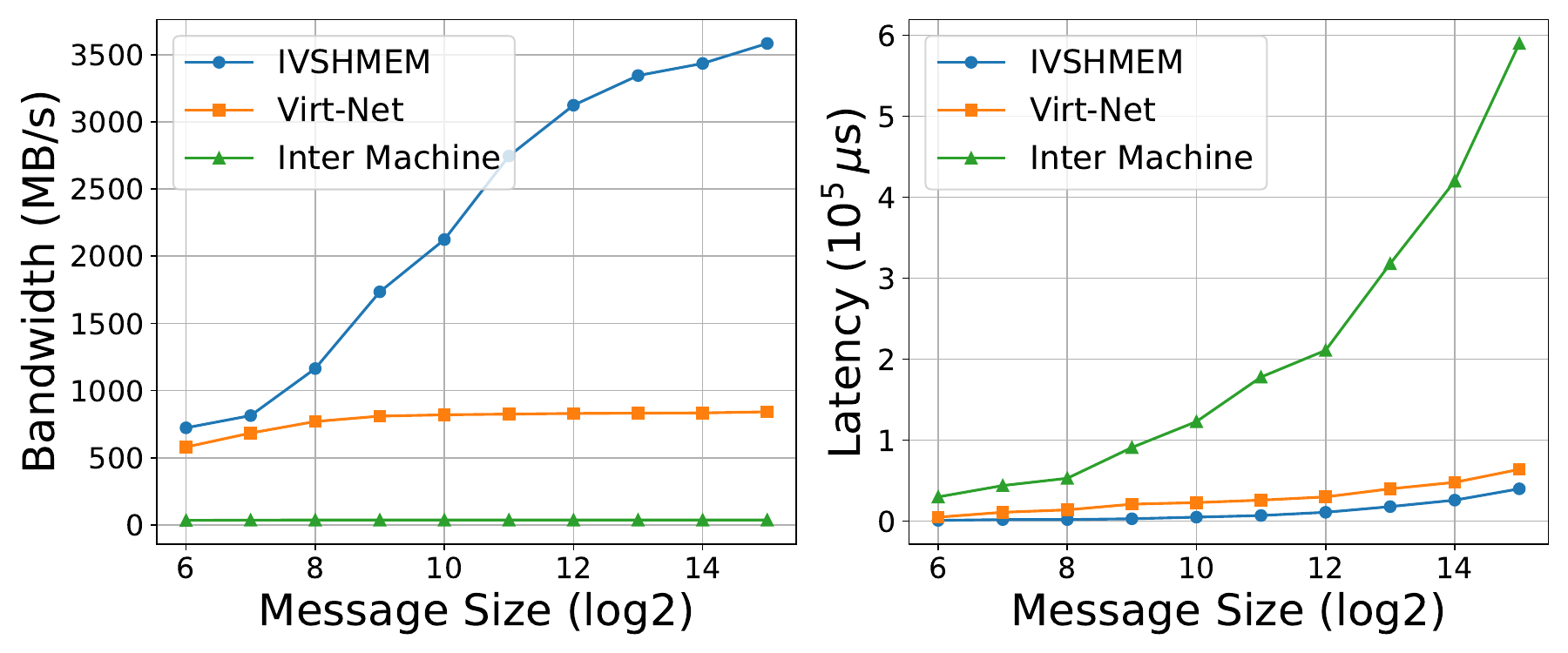}
  \end{center}
  \caption{\label{fig:shm_perf_gain} A performance comparison between IVSHMEM and other inter-VM communication methods. }
\end{figure}

Inter-VM communication in these environments is critical. Traditional
approaches, such as TCP/UDP over a network stack or even UART-based messaging,
often lack sufficient speed and resource efficiency\cite{ren2016shared}.
Alternative solutions like VirtIO offer a para-virtualized communication
mechanism through ring buffers (VirtQueues), but they do not fully leverage the
benefits of shared physical memory.\cite{russell2008virtio} IVSHMEM(Inter-VM
Shared Memory) addresses these limitations by mapping each VM's virtualized PCI
device to a common physical memory region, allowing rapid data exchange through
shared memory\cite{zhang2023optimization,li2023rtism,ren2016shared}. As
illustrated in Figure~\ref{fig:shm_perf_gain}, IVSHMEM achieves substantially
higher bandwidth and lower latency compared to VirtIO-based networking
(Virt-Net) and inter-machine TCP communication over Ethernet, highlighting its
superior performance for inter-VM communication. Despite its performance
advantages, this method introduces significant security challenges; multiple
VMs accessing the same memory space creates vulnerabilities where a compromised
or malicious VM could potentially access or modify data belonging to another
VM\cite{sreenivasamurthy2019sivshm}.

This concern is particularly acute in scenarios where critical systems interact
with less secure environments. For instance, when an RTOS communicates with an
Android-based infotainment VM, there is a tangible risk that a malicious
application within Android might tamper with the shared memory
region\cite{reinhardt2014embedded}. Such tampering could result in attacks
ranging from man-in-the-middle to eavesdropping, ultimately compromising system
stability and safety.

In response to these challenges, we propose a secure protocol designed
specifically for IVSHMEM communication. Our approach introduces robust security
measures on top of the IVSHMEM framework, ensuring data integrity and access
control even in an environment with inherent vulnerabilities. While our
protocol does introduce some performance overhead, we have implemented
techniques to mitigate this impact, ensuring that the overhead remains minimal
relative to the performance gains achieved by shared memory communication.

In this paper, we provide a detailed analysis of the security threats
associated with IVSHMEM, explore the limitations of existing inter-VM
communication methods, and describe our protocol's architecture and mitigation
strategies. Through comprehensive evaluation, we demonstrate that our secure
protocol successfully balances between robust security and the high-performance
demands of IVSHMEM applications—such as those found in modern automotive
systems.

\vspace{3pt}
\section{Background} \label{sec:background}
In this section, we examine the IVSHMEM mechanism, outline its challenges, and
review recent progress.

\subsection{IVSHMEM: Mechanism and Architecture}

\begin{figure}[ht]
  \begin{center}
    \includegraphics[width=0.95\linewidth]{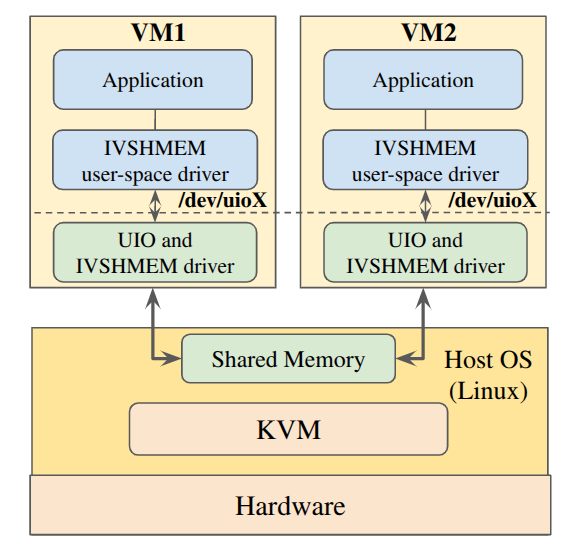}
  \end{center}
  \caption{\label{fig:ivshmem_arch} The overview of IVSHMEM architecture }
\end{figure}

IVSHMEM is a specialized implementation of shared memory IPC designed for
virtualized environments. It emulates a virtual PCI device to expose the shared
memory's base address and size to guest
VMs\cite{qemu_ivshmem_spec,acrn_ivshmem_spec}. As shown in
Figure~\ref{fig:ivshmem_arch}, the hypervisor exposes the shared-memory region
to the guest VM, and the UIO kernel driver maps the emulated PCI device,
allowing applications to access that memory directly via \verb|/dev/uioX|. The
design leverages the standardized PCI configuration to facilitate memory
mapping and efficient communication. Specifically, IVSHMEM utilizes:

\begin{itemize}
  \item \textbf{BAR0 (Base Address Register 0):} This region (256 bytes of MMIO) holds the device registers, which control the operation of the virtual device.
  \item \textbf{BAR1:} It contains the MSI-X table and Pending Bit Array (PBA), primarily used by the IVSHMEM doorbell mechanism for signaling interrupts.
  \item \textbf{BAR2:} This is mapped to the shared memory object, providing a direct communication channel between VMs.
\end{itemize}

The doorbell interrupt mechanism enabled by this configuration allows VMs to
notify one another when new data is available, ensuring efficient core
utilization and reducing latency in inter-VM communication\cite{kazmi2004pci}.

\subsection{Security Concerns of Shared Memory}
Shared-memory IPC within a single OS benefits from well-established
protections, including file access controls, sandboxing, and security modules
such as SELinux or AppArmor. IVSHMEM, however, presents distinct challenges
\cite{spencer1999flask,apparmor_docs}. In a traditional OS environment, the OS
enforces strict access controls over shared memory regions, ensuring that only
authorized processes can read or write data. However, these protections doesn't
work when multiple, potentially untrusted VMs share the same memory space. A
compromised or malicious VM could easily access or tamper with data in the
shared region, leading to unauthorized data disclosure, corruption, or even
system instability\cite{sreenivasamurthy2019sivshm}.

\subsection{Secure Communication Over Insecure Channels}

The challenge of ensuring secure communication in IVSHMEM environments is
analogous to securing communication over the Internet, where multiple parties
exchange information over an inherently insecure channel. In network
communications, protocols such as TLS rely on key exchange mechanisms, mutual
authentication, and end-to-end encryption to safeguard data integrity and
confidentiality\cite{rescorla2018transport,satapathy2016comprehensive}.
Similarly, secure multi-party communication techniques, such as Diffie-Hellman
key exchange and advanced encryption standards, are employed to establish trust
even when the channel is compromised\cite{diffie2022new,rahman2014advanced}.

In the context of IVSHMEM, the situation is even more complex because multiple
services must share the same restricted memory space as a communication
channel. This necessitates designing a secure protocol that ensures
confidentiality and integrity, similar to TLS or other network security
protocols, while also accommodating the shared nature of the memory resource.
Our research addresses these challenges by proposing a secure protocol that
protect the data transmitted via IVSHMEM, while also mitigating the performance
overhead typically associated with such security measures.

\subsection{Recent Progress of IVSHMEM Communication}

Recent research and developments in IVSHMEM communication have focused on
balancing performance and security, with various approaches having distinct
trade-offs.

The SIVSHM project introduces a segmentation approach to IVSHMEM, enhancing
security by isolating shared memory regions among VMs. However, this strict
isolation introduces notable overheads from reduced buffer size.
\cite{sreenivasamurthy2019sivshm}.

Performance-centric solutions, such as XenLoop and MemPipe, have been proposed
to optimize IVSHMEM by improving transparency and reducing
latency\cite{zhang2007xensocket}\cite{ren2016shared}\cite{zhang2016workload}\cite{wang2008xenloop}\cite{diakhate2008efficient}\cite{macdonell2011shared}.
These solutions integrate seamlessly with traditional socket-based network
stacks, allowing applications to benefit from high-speed shared memory
communication without explicit changes. However, these solutions primarily
prioritize performance and transparency and lack of mechanisms for security
threats\cite{ren2016shared}.

Other approaches have leveraged hypervisor-managed policies and features, like
Xen's grant tables, to enforce finer-grained security controls. Grant tables
establish explicit, controlled memory-sharing agreements between VMs,
restricting access to designated regions. However, this technique is tied to a
specific hypervisor and does not provide a general-purpose driver interface
\cite{xenGrantTables}.

Overall, while significant progress has been made in enhancing both the
performance and security of IVSHMEM, current solutions optimize for one at the
expense of the other. We aim to design a Secure IVSHMEM protocol that delivers
security features while keeping performance overhead to a minimum.

\vspace{3pt}
\section{Threat Model}\label{sec:threat_model}

In this section we define the assets to be protected, the adversary's
capabilities, our trust assumptions, concrete threat scenarios with
corresponding defenses, the security goals achieved, and known limitations.

\subsection{Assets}
\begin{itemize}
  \item \textbf{Shared-Memory Contents:} All plaintext data exchanged via the IVSHMEM region (e.g., sensor readings, control commands).
  \item \textbf{VM Identities:} Certificates and private keys provisioned to each VM by the hypervisor CA.
\end{itemize}

\subsection{Adversary Model}
We consider an attacker with the following capabilities and goals:
\begin{description}
  \item \textbf{Location:} Co-resident on the same host, either in another VM or with limited
        host privileges.
  \item \textbf{Privileges in Guest:} May be an unprivileged process or even gain root in one VM, and thus can open and attempt to \texttt{mmap()} the IVSHMEM region directly.
  \item \textbf{Goals:}
        \begin{enumerate}
          \item \emph{Confidentiality breach:} Read plaintext data from another VM's IVSHMEM region.
          \item \emph{Integrity breach:} Inject or tamper with messages in the shared region.
          \item \emph{Authentication breach:} Impersonate a VM by forging or replaying handshake messages.
        \end{enumerate}
\end{description}

\subsection{Trust Assumptions}
\begin{itemize}
  \item \textbf{Trusted Hypervisor:} Each VM trusts the hypervisor as the root of trust; although VMs do not inherently trust the IVSHMEM communication channel, they rely on the hypervisor acting as a Certificate Authority (CA) to issue, sign, and validate VM certificates.
  \item \textbf{Kernel-Module Enforcement:} All VMs in the system load and execute the same IVSHMEM enforcement kernel module, ensuring uniform, in-kernel access control and preventing any unauthorized memory mappings across the entire platform.
\end{itemize}

\subsection{Security Goals}
Under the above model and countermeasures, our protocol achieves:
\begin{enumerate}
  \item \textbf{Confidentiality:} No application in VM can read another's channel's plaintext data.
  \item \textbf{Integrity:} Any tampering with shared data is detected by authentication tags.
  \item \textbf{Mutual Authentication:} Only application in VMs with valid, hypervisor-signed certificates complete the handshake.
\end{enumerate}

\subsection{Limitations of Conventional Security Protocols}

Conventional end-to-end security protocols such as TLS, IPsec, or DTLS are
ill-suited to the IVSHMEM use-case for several reasons:

\begin{enumerate}
  \item \textbf{Performance Overhead:}
        Conventional TLS requires symmetric encryption and decryption on each record, which breaks IVSHMEM's zero-copy path and forces additional data copies and context switches\cite{rescorla2018transport}\cite{muller2014benchmarking}.  In a high-throughput IVSHMEM environment—where direct page mappings sustain multiple gigabytes per second—this per-record crypto overhead introduces unacceptable latency and CPU load.

  \item \textbf{Limited Shared-Memory Capacity:}
        Unlike network channels, IVSHMEM regions are fixed and small (e.g., 1 MiB)\cite{qemu_ivshmem_spec}\cite{acrn_ivshmem_spec}. To prevent a malicious VM or service from overwhelming the shared-memory resource, the hypervisor must strictly assign and enforce per-service channel quotas. Conventional socket-based protocols provide no mechanism for hypervisor-driven, size-limited region allocation or fine-grained resource control.

  \item \textbf{Inadequate Fit for End-to-End Schemes:}
        Conventional end-to-end protocols (e.g., IPsec, SSH, DTLS) assume a network stack with IP addresses, ports, and hostnames or DNS names to establish and authenticate channels. IVSHMEM operates entirely in-host via PCI BAR mappings without any network identifiers, so these protocols cannot provide true end-to-end security for shared-memory communication or integrate with hypervisor-managed channel assignment.

\end{enumerate}

Conventional end-to-end protocols such as TLS, IPsec, and DTLS are ill-suited
for IVSHMEM communication because they (1) impose per-record cryptographic
overhead that breaks zero-copy
performance\cite{shue2005analysis,muller2014benchmarking}, (2) offer no
mechanism to enforce fixed, size-limited shared-memory quotas, and (3) depend
on network-layer identifiers (e.g., hostnames, IP addresses) while lacking
support for hypervisor-driven channel
assignment\cite{satapathy2016comprehensive}. Securing IVSHMEM therefore
requires a specialized protocol that leverages IVSHMEM's in-host, fixed-region
semantics and hypervisor control, providing end-to-end protection with
minimizing additional performance overhead.

\vspace{3pt}
\section{Design Proposal}\label{sec:design_proposal}

In this section, we present the Secure IVSHMEM design, which aims to provide
dedicated, zero-copy shared-memory channels between VM services while
preventing unauthorized access, spoofing, and impersonation. Our approach
combines three complementary mechanisms: \emph{service-based channel
  separation} to allocate isolated IVSHMEM regions per service pair,
\emph{granular kernel-module enforcement} to block any unauthorized
\texttt{open}, \texttt{mmap}, or I/O operations, and a
\emph{hypervisor-mediated mutual-authentication handshake} to establish trust
on channel setup. Together, these components ensure end-to-end security with
negligible impact on IVSHMEM's high-performance communication.

\subsection{Service-based Channel Separation}

\begin{figure*}[ht]
  \begin{center}
    \includegraphics[width=0.9\linewidth]{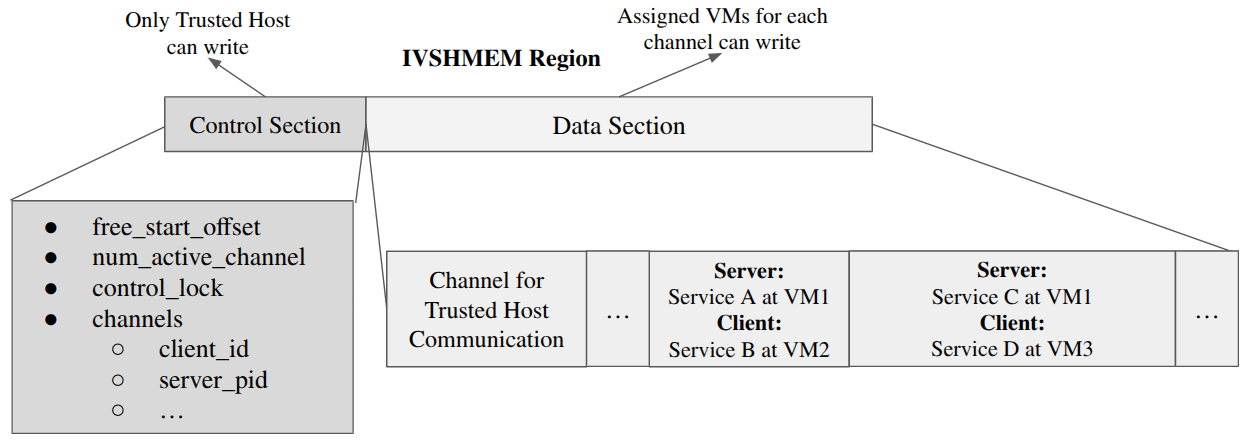}
  \end{center}
  \caption{ Service-based channel separation in Secure IVSHMEM: the trusted host controls metadata in the Control Section and is assigned its own channel in the Data Section for host–VM communication, while each service pair communicates over its dedicated Data Section channel.}

  \label{fig:channel_separation}
\end{figure*}

As illustrated in Figure~\ref{fig:channel_separation}, our design divides the
IVSHMEM architecture into two primary sections: the \textbf{Control Section}
and the \textbf{Data Section}. The Control Section is a fixed-size region where
a trusted host stores dynamic configurations related to data allocation, while
the Data Section is where the service in virtual machines (VMs) actually read
and write data through their assigned channels. Importantly, only the trusted
host has permission to modify data in the Control Section, and each VM is
restricted to reading and writing only to its designated channels within the
Data Section.

The \textbf{Data Section} comprises multiple channels, with each channel
serving as a dedicated buffer space for a specific server and client service
pair. For each pair, a dedicated channel is allocated, and the Control Section
dynamically adjusts its size based on the activation of channels.

For example, consider a scenario where Service A in VM1 needs to send data to
Service B in VM2. In this case, the trusted host allocates an initial channel
with a buffer size of 512 KiB. The control information for this allocation is
written into the Control Section, and only Service A and Service B are
permitted to access the channel's buffer. Additionally, the size of the channel
buffer can be adjusted based on the usage patterns between the services.

An exception to this rule is the first channel in the Data Section. This
channel is of a fixed-size and is exclusively used for communication between
the trusted host and the VMs, such as during the initial handshake when a VM
sends data to the trusted host. All VMs have access to this channel.

The \textbf{Control Section} maintains all of the metadata needed for buffer
allocation and channel management. It tracks the next free offset, the number
of active channels, and protects updates with a lock. Per-channel metadata
(service IDs, process IDs, buffer addresses and sizes) is stored in an internal
array. Channels use this information to coordinate reads and writes to their
assigned regions.

\subsection{Granular Kernel Module Enforcement}

To enhance the security of the IVSHMEM framework, we propose a granular access
control mechanism that restricts access to the shared memory channels on a
per-application basis. This mechanism is implemented via a dedicated kernel
module that operates on top of the IVSHMEM device driver.

\subsubsection{Kernel Module Integration}

Our kernel module hooks all IVSHMEM-related system calls, including
\texttt{open}, \texttt{read}, \texttt{write}, and \texttt{mmap}, as well as any
I/O control operations targeting the IVSHMEM device. On each intercepted call,
the module retrieves the caller's \texttt{service\_id} and checks the Control
Section's metadata to verify that this service identifier matches the channel
being accessed. If the \texttt{service\_id} does not correspond to that
channel's assigned service, the module denies the operation. This enforcement
ensures that only the authorized host or VM service can interact with its
designated shared-memory region.

\subsubsection{Channel-Specific Enforcement}

Each channel within the Data Section is allocated to a specific pair of
services (e.g., a server and a client). The kernel module uses the control
section's metadata to determine channel assignments and enforces strict access
control, permitting operations only on the designated channel buffers.

\subsection{Hypervisor-Mediated Mutual-Authentication Handshake}

A secure, hypervisor-mediated handshake is essential for IVSHMEM because it
protects against spoofing and impersonation on both sides of the shared-memory
channel. In our model, the trusted host (e.g., dom0 in Xen, SOS in
ACRN\cite{li2019acrn}, or the host in QEMU/KVM) cannot simply trust that any
service presenting a request truly owns its claimed endpoint; similarly, a VM
service cannot blindly accept control messages or credentials from the host
without risk of impersonation. By embedding a mutual-authentication handshake
into the Control Section—that is, having each party present and verify
cryptographic credentials tied to its service id—the host first confirms that
the requesting service is one of its pre-registered, trusted entities, and then
the VM service validates that the host's response really comes from the genuine
hypervisor authority. Only once both directions of identity proof succeed do we
allocate a dedicated, zero-copy channel in the Data Section. This two-way
validation thwarts malicious actors on either side and ensures end-to-end trust
before any IVSHMEM communication occurs.

Our proposed protocol is different from the conventional internet based
security protocol in that 1) The trusted host (hypervisor) is not merely a
passive participant but is responsible for allocating finite resources and
establishing the communication channel and 2) The hypervisor functions as a
certification authority (CA)\cite{perez2006vtpm}, validating service
credentials and orchestrating the creation of dedicated secure channels between
clients and servers.

The detailed handshake protocol steps are provided below, demonstrating the
mutual authentication processes that lead to the establishment of a
confidential IVSHMEM channel.

\begin{figure}[ht]
  \begin{center}
    \includegraphics[width=1.0\linewidth]{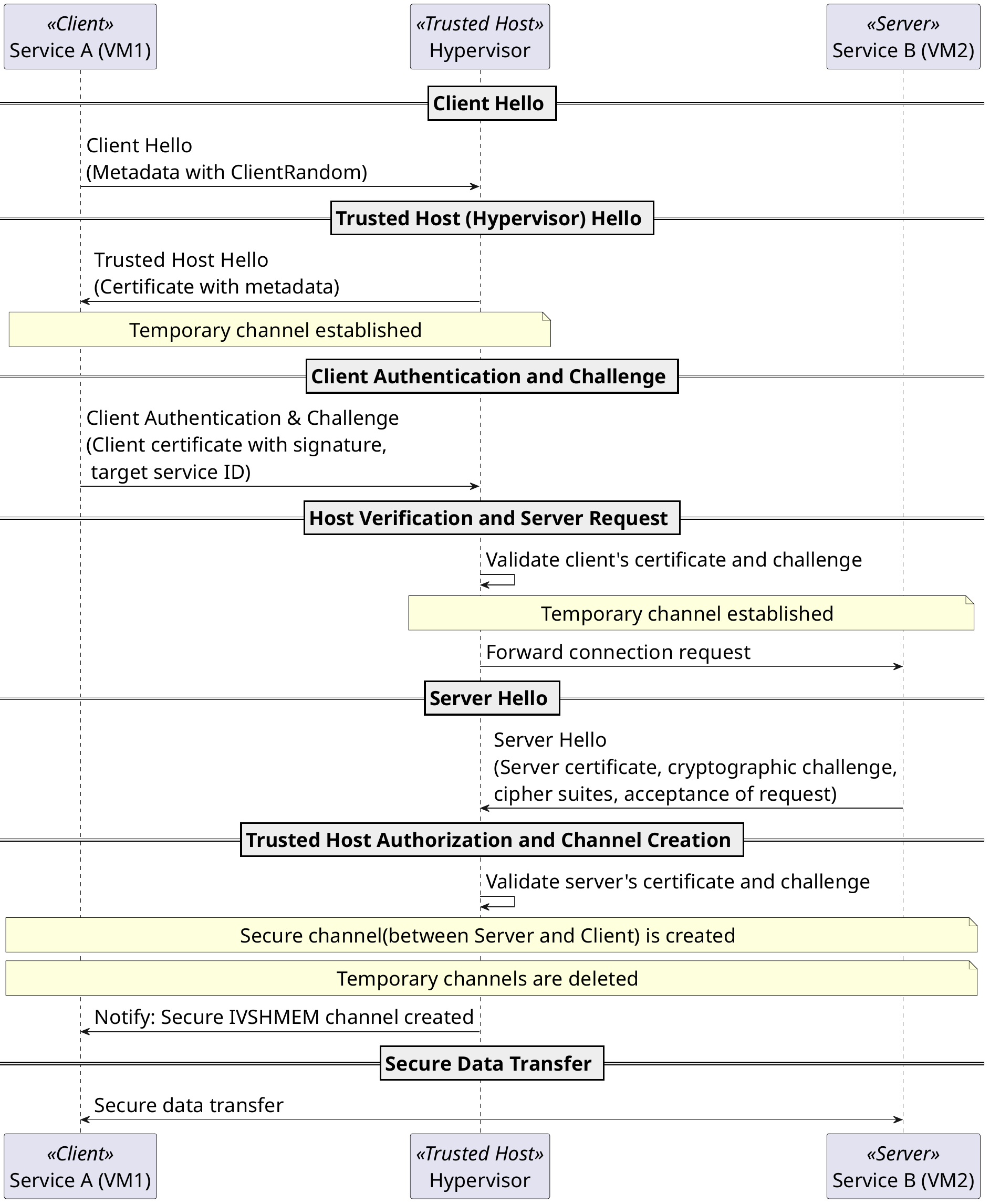}
  \end{center}
  \caption{ Secure IVSHMEM handshake flow: eight steps from Client Hello through Host authorization to establishment of the dedicated shared-memory channel }
  \label{fig:handshake_protocol}
\end{figure}

\vspace{3pt}
\subsubsection{Protocol Steps (Fig.~\ref{fig:handshake_protocol})}

\begin{enumerate}[label=\arabic*.]
  \item \textbf{Client Hello:}
        \begin{itemize}
          \item \textbf{Purpose:} Initiate the handshake and propose communication parameters.
          \item \textbf{Message Contents:} protocol version \& extensions, supported cipher suites, client identity (service ID, PID, VM ID), nonce + timestamp for replay protection.
        \end{itemize}

  \item \textbf{Trusted Host Hello:}
        \begin{itemize}
          \item \textbf{Purpose:} Acknowledge the client's request and provide trusted credentials.
          \item \textbf{Message Contents:} host certificate (± selected ciphers);
          \item \textbf{Action:} A temporary secure channel is created between the client and the trusted host.
        \end{itemize}

  \item \textbf{Client Authentication and Challenge:}
        \begin{itemize}
          \item \textbf{Purpose:} Enable explicit mutual authentication.
          \item \textbf{Message Contents:} The client certificate, signature over nonce, target server's service ID.
        \end{itemize}

  \item \textbf{Host Verification and Server Request:}
        \begin{itemize}
          \item \textbf{Purpose:} Verify the client's credentials and initiate communication with the server.
          \item \textbf{Action:}
                \begin{itemize}
                  \item The trusted host validates the client's certificate and challenge.
                  \item Upon successful validation, the trusted host creates a temporary channel for
                        the server and forwards a secure connection request to the server, including
                        the nonce.
                \end{itemize}
        \end{itemize}

  \item \textbf{Server Hello:}
        \begin{itemize}
          \item \textbf{Purpose:} Server passes its credentials and decision to accept the client's request.
          \item \textbf{Message Contents:} server certificate, signature challenge, supported ciphers, acceptance flag.
        \end{itemize}

  \item \textbf{Trusted Host Authorization and Channel Creation:}
        \begin{itemize}
          \item \textbf{Purpose:} Establish a secure, dedicated IVSHMEM channel for data transfer between the server and client.
          \item \textbf{Action:}
                \begin{itemize}
                  \item The trusted host verifies the server's certificate and the corresponding
                        challenge.
                  \item The trusted host creates a communication channel for the server and client and
                        notifies the client.
                  \item The trusted host deletes temporary channels previously established with both
                        the client and the server.
                \end{itemize}
        \end{itemize}

  \item \textbf{Secure Data Transfer:}
        \begin{itemize}
          \item \textbf{Purpose:} Enable protected communication.
          \item \textbf{Action:} The client and server commence secure data transfer over the authorized IVSHMEM channel.
        \end{itemize}

  \item \textbf{Session Management (Optional Enhancements):}
        Implement mechanisms (similar to TLS session tickets) for efficient session
        resumption without repeating the full handshake process.
\end{enumerate}

\begin{table*}[h]
  \centering
  \begin{tabular}{p{4cm} p{7cm} p{6cm}}
    \toprule
    \textbf{Scenario}                                                                                                                               & \textbf{Attack Flow} & \textbf{Countermeasure} \\
    \midrule
    Eavesdropping \& Unauthorized Mapping                                                                                                           &
    Rogue VM or process calls \texttt{open}/\texttt{mmap} on IVSHMEM and reads raw data from an unassigned channel before authentication.           &
    Kernel module enforces per-channel access control, blocking any \texttt{open}, \texttt{mmap}, or I/O until the channel is marked \texttt{ESTABLISHED} by the handshake.                          \\
    \addlinespace
    Replay \& Man-in-the-Middle Attacks                                                                                                             &
    Adversary captures or intercepts handshake messages (certificates, nonces) and replays or tampers with them to hijack or impersonate a service. &
    Mutual-authentication handshake with CA-signed certificates and nonce-based challenge prevents replay and ensures both endpoints verify peer identity.                                           \\
    \bottomrule
  \end{tabular}
  \caption{Key threat scenarios and corresponding Secure IVSHMEM defenses}
  \label{tab:threat-defenses-brief}
\end{table*}

\subsection{Ring Buffer and Anticipation Timing Window}

\begin{figure}[!h]
  \centering
  \includegraphics[width=0.35\textwidth]{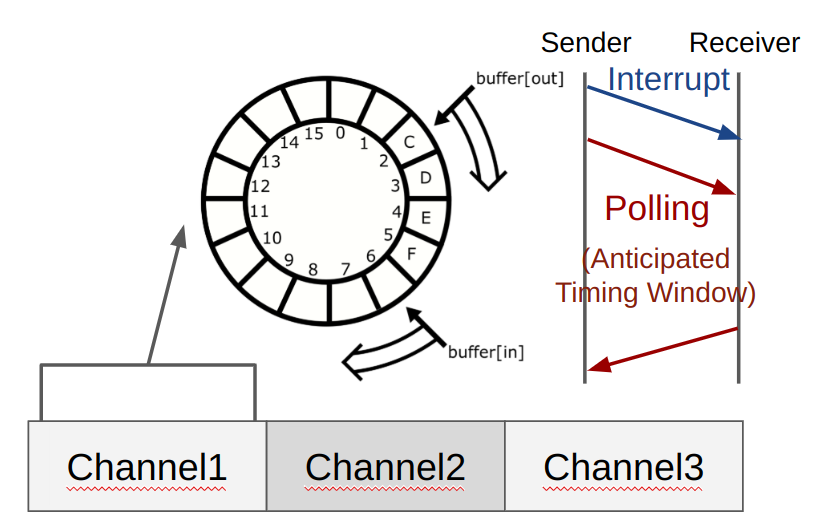}
  \caption{ The IVSHMEM protocol is optimized using a Ring Buffer and an Anticipation Time Window mechanism.}
  \label{fig:ringbuf_polling}
\end{figure}

To enhance performance in our IVSHMEM protocol design, we adopt two key
mechanisms: a ring buffer and an anticipation timing window. Ring buffers are
widely used in shared memory communication due to their low-overhead, lock-free
design, which supports efficient producer-consumer patterns. This structure
allows continuous data exchange without the need for frequent synchronization,
thereby improving throughput and reducing contention.

In addition, we incorporate the anticipation timing window mechanism,
originally proposed in \cite{zhang2016workload}, to mitigate the overhead
associated with frequent interrupt handling. Rather than triggering an
interrupt (IRQ) for every data transmission, which incurs significant
system-wide context switching overhead, the receiver VM periodically polls the
shared buffer for a predefined time window. This approach reduce interrupt cost
while maintaining responsiveness, ultimately reducing communication latency and
improving overall efficiency.

The performance benefits of these optimizations are evaluated in detail in
Section~\ref{sec:measurements}.


\vspace{3pt}

\section{Implementation}

In this section, we implment Secure IVSHMEM through three components: a kernel
module that hooks into the UIO PCI driver\cite{weisbach2011generic} to enforce
per-channel access control, a user-space OpenSSL-based mutual-authentication
handshake over the control page, and a BSD-socket-style library for zero-copy
ring-buffer data transfers on authenticated channels.

\subsection{Kernel-Module Integration via Dynamic Hooks}

We build on top of the existing UIO PCI driver for IVSHMEM by inserting a
lightweight kernel module that intercepts system calls(\texttt{mmap()},
\texttt{read()} and \texttt{write()})to enforce per-channel access control.

\paragraph{Hook Implementation}
Using a combination of kprobes and ftrace\cite{gebai2018survey}, our module
attaches to the IVSHMEM driver's \texttt{mmap()} entry point. In the hook we:

\begin{itemize}
  \item Extract the \texttt{vma} pointer from the CPU registers.
  \item Compute the requested mapping's channel ID and range.
  \item Look up the authorized-PID list stored in the IVSHMEM control section.
  \item If \texttt{current->pid} is not present or the channel is not marked
        \texttt{AUTHORIZED}, force-return \texttt{-EPERM} and skip the real handler.
\end{itemize}

\paragraph{Policy Management}
An in-kernel hash table of \texttt{policy\_entry} structs—keyed by
channel\_id—tracks which PIDs are permitted each channel. The hypervisor
populates this table at boot, and upon handshake completion a simple ioctl
marks the corresponding entry as \texttt{AUTHORIZED}.

\paragraph{Cleanup}
On module unload or VM teardown, all probes are unregistered and the policy
table cleared, restoring the original UIO driver behavior.

This dynamic-hook approach adds minimal overhead, preserves zero-copy data
mappings for authorized clients, and requires no modification to the upstream
IVSHMEM driver.

\subsection{Handshake Implementation}

\begin{figure}[!h]
  \centering
  \includegraphics[width=0.5\textwidth]{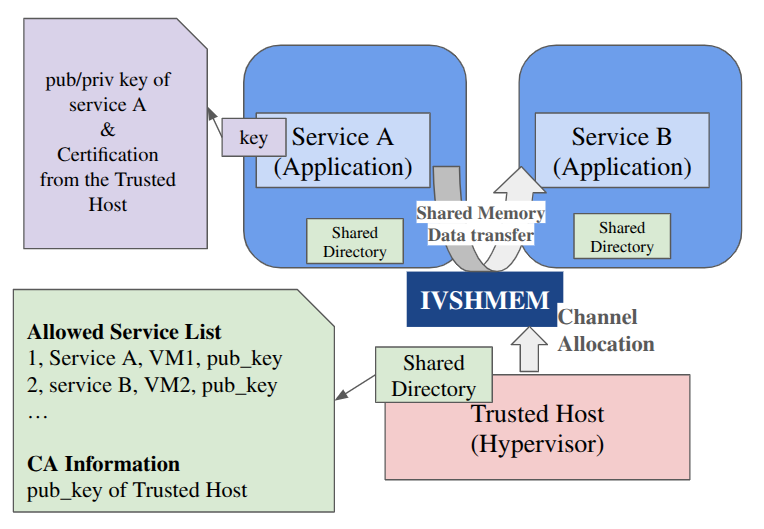}
  \caption{ Credential management: the hypervisor publishes a CA certificate and per-service public keys in the Allowed Service List for Secure IVSHMEM }
  \label{fig:handshake_credential}
\end{figure}


As illustrated in Figure~\ref{fig:handshake_credential}, the
mutual-authentication handshake is implemented entirely in user-land using
OpenSSL\cite{viega2002network} and leverages a IVSHMEM as an in-host transport.
During the initial setup, the hypervisor generates a 4096-bit RSA CA key and
self-signed certificate. It then builds an \emph{allowed-service list} that
maps each (service\_ID, VM\_ID) pair to its corresponding public key, and
publishes that list together with the CA's public certificate to all VMs via a
shared directory. Next, the hypervisor issues 2048-bit RSA key pairs for each
registered service, retaining the public key in its allowed-service list and
provisioning the private key to the guest VM. Each service in the VM uses this
private key to generate its certificate when it performs the handshake. At
runtime, host and VM exchange certificates over the IVSHMEM control channel,
verify each peer's certificate against the CA and the allowed-service list, and
send signed acknowledgments to confirm mutual credential validation. Once both
sides have verified the signature, the VM mark the kernel module that the
channel is \texttt{ESTABLISHED}, enabling zero-copy operations under the
authenticated session.

\subsection{Application Library Implementation}

To simplify adoption of Secure IVSHMEM, we provide a lightweight user-land
library with APIs analogous to BSD sockets:

\begin{itemize}
  \item \texttt{ivshmem\_listen(vm\_id, service\_id)} – being ready to accept connections from target on a dedicated IVSHMEM channel
  \item \texttt{ivshmem\_connect(vm\_id, service\_id)} – initiate a connection to the target, performing the handshake over the control page
  \item \texttt{ivshmem\_send(buf, len)} – copy \texttt{len} bytes from \texttt{buf} into the ring buffer slots of the established channel and ring the doorbell
  \item \texttt{ivshmem\_receive(buf, len)} – poll the ring buffer for new data, copy up to \texttt{len} bytes into \texttt{buf}
\end{itemize}

Under the hood, \texttt{ivshmem\_listen} and \texttt{ivshmem\_connect} map the
\texttt{(vm\_id, service\_id)} tuple to a hypervisor-assigned PCI BAR channel,
then carry out the certificate exchange and mutual validation over the IVSHMEM
control region.

For payload transfers, the library allocates a ring buffer within the IVSHMEM
data section so that producers and consumers operate without blocking:

1. The sender writes into its next available slot in the ring buffer and triggers a doorbell interrupt.
2. The receiver, polling the doorbell and head pointer, copies the data into its local buffer and advances the consumer index.
3. A secondary doorbell notifies the sender that the slot is free.

By combining zero-copy ring buffers with doorbell interrupts followed by brief
polling, this API achieves near-native IVSHMEM performance while enforcing
end-to-end authentication. We evaluate its throughput in
Section~\ref{sec:measurements}.


\section{Measurements} \label{sec:measurements}

The experiments were run on two Linux guest VMs, each using the 6.12.10-0-lts
kernel and provisioned with 2 GiB of RAM and 4 vCPUs. The host is an x86\_64
machine with an Intel® Core™ Ultra 7 155H (VT-x enabled, 400 MHz-4.8 GHz) and
32 GiB of DDR memory. To minimize interference, each VM was pinned to its own
physical cores, and both the control channel and IVSHMEM devices were
instantiated via QEMU's full-virtualized interfaces

\subsection{Latency Overhead for Initial Handshake}

We record the one-time handshake cost—from the initial
\texttt{ivshmem\_connect()} call through certificate exchange and verification
until the first confirmation—and then measure steady-state data-plane
round-trip latency for each 32-bit write (plus doorbell notification),
comparing vanilla IVSHMEM to Secure IVSHMEM.

\begin{table}[h]
  \centering
  \small
  \begin{tabular}{lcc}
    \toprule
    \textbf{Operation}  & \textbf{Vanilla (µs)} & \textbf{Secure (µs)} \\
    \midrule
    Initial Handshake   & -                     & 90\phantom{.0}       \\
    Round-Trip Transfer & 8.1                   & 8.4                  \\
    \bottomrule
  \end{tabular}
  \caption{Round-trip latency for a single 32-bit integer.  Secure IVSHMEM incurs a 90µs handshake cost, but per-message latency afterward is within 5 \% of the vanilla baseline.}
  \label{tab:latency}
\end{table}

As shown in Table~\ref{tab:latency}, although the initial handshake adds a
modest 90µs one-time overhead, the steady-state data-plane latency (8.4 µs) is
almost identical to vanilla IVSHMEM (8.1 µs). This shows that, although the
initial handshake introduces some latency, our security mechanism adds
negligible per-message overhead once the session is established.

\subsection{Kernel Module Enforcement Overhead}

For each experiment, a total of 32\,GiB of random data was transferred to
evaluate raw throughput and quantify the overhead introduced by granular access
control via a kernel module.

\begin{figure}[!h]
  \centering
  \includegraphics[width=\linewidth]{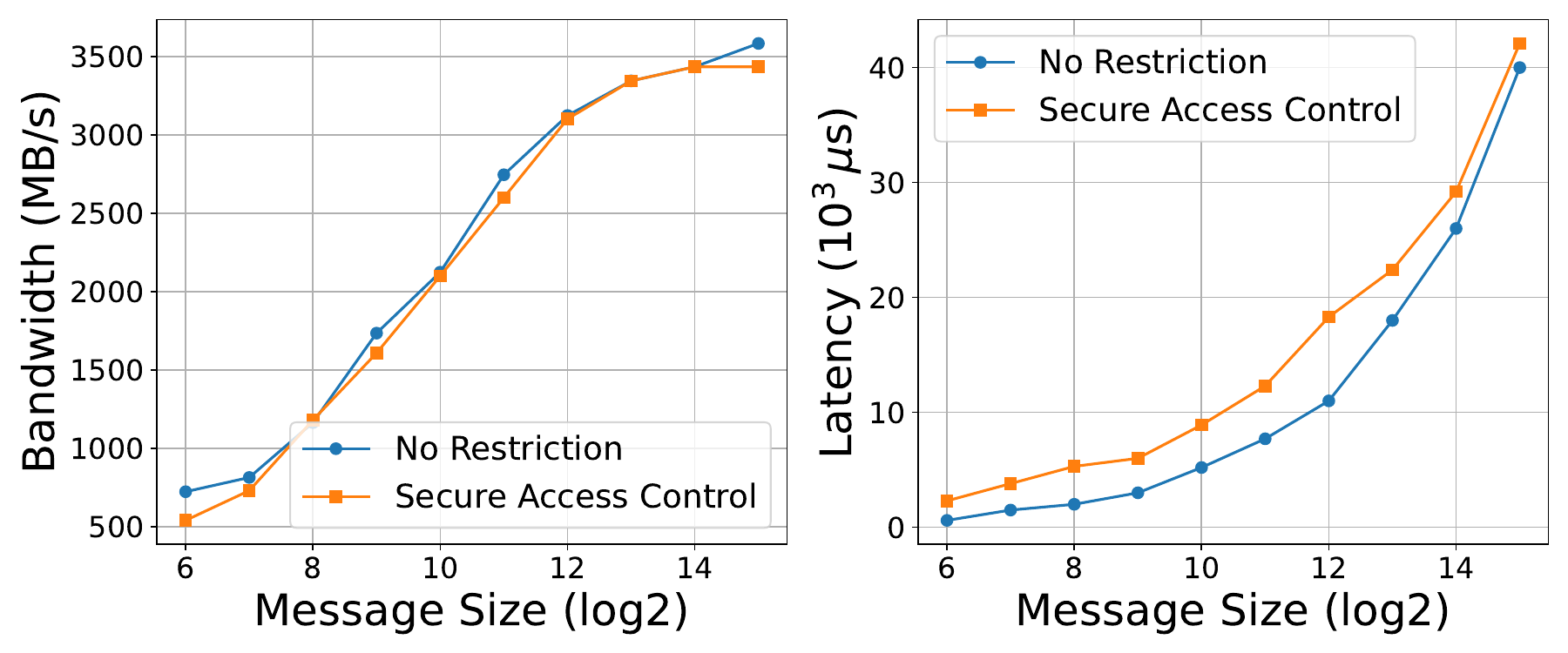}
  \caption{Bandwidth comparison of Secure IVSHMEM Protocol versus Vanilla IVSHMEM.}
  \label{fig:secure_ivshmem_perf}
\end{figure}

Figure~\ref{fig:secure_ivshmem_perf} compares the performance of IVSHMEM with
and without access control, across message sizes ranging from $2^6$ to $2^{15}$
bytes. For small messages (<= $2^8$\,B), the integration of kernel-module hooks
incurs a throughput reduction of approximately 20--25\%. However, this overhead
diminishes rapidly with increasing message size. For transfers >= 1\,KiB, the
bandwidth of Secure IVSHMEM remains within 5\% of the unmodified baseline,
indicating that the additional hooking logic imposes minimal overhead at larger
scales.

In terms of latency, the kernel-module integration introduces an overhead of up
to 25--30\% for small messages, but this impact also decreases as the message
size grows. For messages $\geq$ 1\,KiB, the latency overhead drops to below
10\%, demonstrating that the cost of access control becomes negligible for
larger payloads.

\subsection{Performance Optimization}

\begin{figure}[!h]
  \centering
  \includegraphics[width=\linewidth]{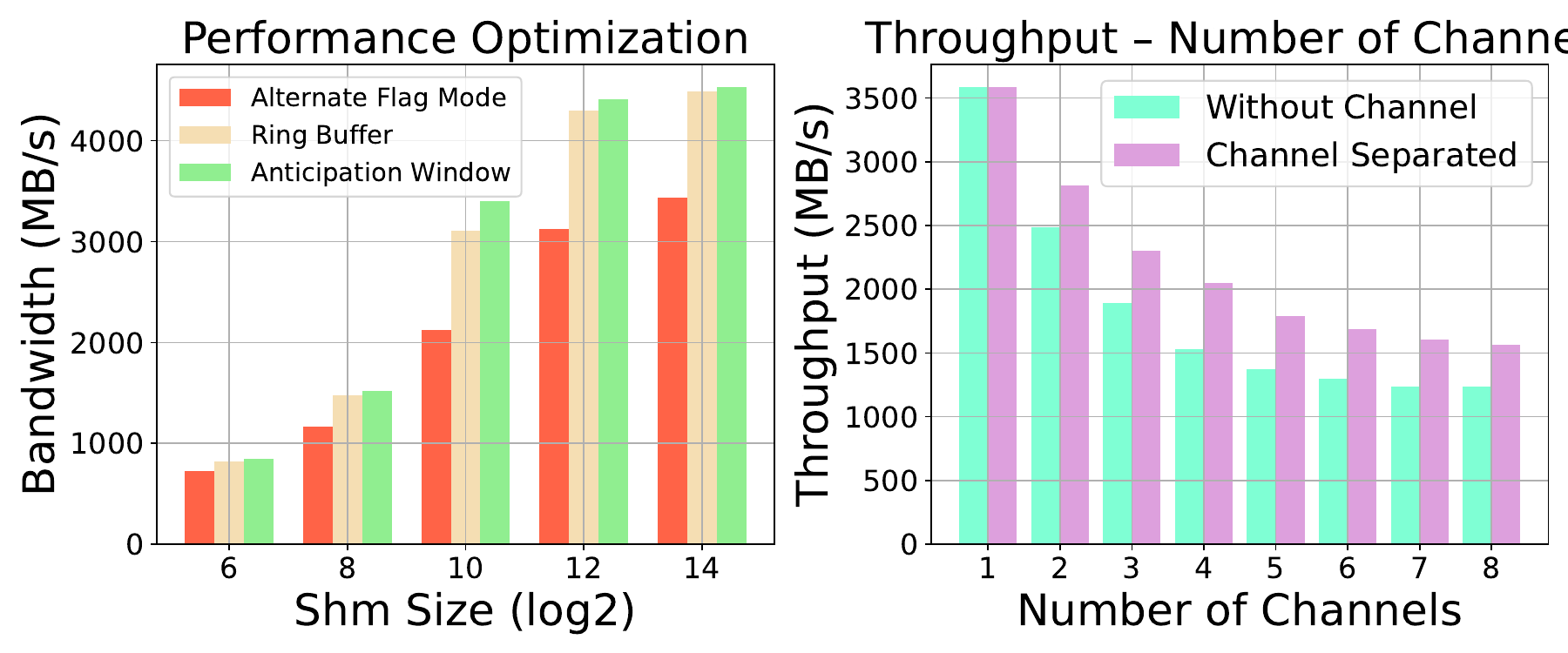}
  \caption{
    Performance improvements achieved through the use of a ring buffer and anticipation timing window. Channel separation further enhances bandwidth in multi-producer, multi-consumer scenarios
  }
  \label{fig:shm_ablation}
\end{figure}

Figure~\ref{fig:shm_ablation} compares the performance of three IVSHMEM
communication designs across increasing shared memory sizes: a basic
message-passing scheme, a ring-buffer implementation, and a design enhanced by
an anticipation timing window.

The results demonstrate substantial performance gains from both optimizations.
Applying the ring-buffer architecture—which leverages batching and zero-copy
techniques—reduces system call and memory copy overhead and increases bandwidth
by more than 30\% once the shared-memory size reaches \(\geq 2^{10}\) bytes,
with an improvement of approximately 46\% at \(2^{10}\) and sustained gains
above 30\% at larger sizes. The anticipation timing window further improves
throughput by minimizing the cost of frequent interrupt handling, delivering
performance boosts of at least 4\% across all tested sizes. Together, these
enhancements enable more efficient use of shared memory and higher
communication bandwidth as the message size increases.

Additionally, one of the interesting things is that channel separation brings
additional performance gains in multi-producer, multi-consumer scenarios.
Figure~\ref{fig:shm_ablation} depicts throughput as a function of the number of
concurrent producer/consumer pairs (separate IVSHMEM channels). With a single
channel, both vanila and Secure IVSHMEM sustain
\(\sim3.58\,\mathrm{GiB}/\mathrm{s}\). As the channel count increases to eight,
per-channel throughput decreases to \(\sim1.56\,\mathrm{GiB}/\mathrm{s}\) for
vanilla and \(\sim1.23\,\mathrm{GiB}/\mathrm{s}\) for the secure variant. This
performance gain stems from reduced lock contention and parallel, lock-free
processing across channels.

In summary, although our Secure IVSHMEM protocol incurs a one-time handshake
latency when establishing each channel, it introduces negligible per-message
overhead in both latency and bandwidth once the channel is established;
moreover, channel separation combined with a ring-buffer and anticipation
timing window delivers additional scalability and throughput gains in
multi-producer/multi-consumer scenarios by reducing lock contention.

\subsection{Security Validation Experiments}

To validate that our kernel-module enforcement reliably blocks unauthorized
access, we designed three attack scenarios reflecting realistic bypass
attempts. Each scenario was exercised 30 times on our testbed with the Secure
IVSHMEM module active:

\begin{description}[leftmargin=0.5cm]
  \item[\textbf{Out-of-Bounds Data Access}]
        \emph{Attack:} \texttt{mmap()} requests a data range that lies outside the limits recorded in the control-section metadata.
  \item[\textbf{Control-Section Access Violation}]
        \emph{Attack:} attempt to read via syscall from the read-only control section.
  \item[\textbf{Impersonation Attack}]
        \emph{Attack:} handshake request using an invalid or replayed credential (wrong service ID or stale nonce).
\end{description}

In all three cases, the kernel module returned \texttt{-EPERM} and no pages or
credentials were granted. A valid data-section mapping (exactly matching the
bounds in control metadata) succeeded and was cleanly unmapped.

\begin{table}[!h]
  \centering
  \small
  \begin{tabular}{l c c}
    \toprule
    \textbf{Test Scenario}           & \textbf{Attempts} & \textbf{Blocked (\%)} \\
    \midrule
    Out-of-Bounds Data Access        & 30                & 30 (100\%)            \\
    Control-Section Access Violation & 30                & 30 (100\%)            \\
    Impersonation Attack             & 30                & 30 (100\%)            \\
    \bottomrule
  \end{tabular}
  \caption{Invalid Access Test Results: all invalid attempts were correctly blocked.}
  \label{tab:sec_results}
\end{table}

\vspace{3pt}
\section{Discussion}\label{sec:discussion}

\paragraph{Hypervisor Independence}
Our Secure IVSHMEM protocol and its kernel-module integration are
hypervisor-agnostic. Both the handshake mechanism and the IVSHMEM driver can be
deployed on any virtualization platform that supports UIO and the IVSHMEM
device, including ACRN and Xen.

\paragraph{Dynamic Channel Buffer Allocation}
In our current prototype, each channel's buffer is allocated as a single
contiguous region for simplicity. To reduce external fragmentation and improve
memory utilization, a page-based or scatter/gather buffer allocation scheme
could be adopted.

\paragraph{Key Exchange and Symmetric Encryption}
While we focus here on authentication and integrity, confidentiality could be
added via symmetric encryption. A TLS-style key-exchange (for example,
ephemeral Diffie-Hellman over the control channel) would introduce only a
modest one-time handshake delay and encryption overhead sacrificing zero-copy
data-plane performance.

\paragraph{Transparency}
Our protocol and kernel module require applications to link against the
IVSHMEM-specific library and invoke its API, rather than using standard socket
or networking calls. Achieving full
transparency\cite{wang2008xenloop,zhang2016workload,zhang2007xensocket,wang2008xenloop}—so
that unmodified applications could communicate over IVSHMEM as if it were
TCP/IP or any other network transport—is beyond the scope of this work and is
left for future exploration (e.g., via socket-API interposition or
hypervisor-level redirection).

\vspace{3pt}
\section{Conclusion}

We have presented Secure IVSHMEM, a protocol that delivers end-to-end
authentication and integrity for in-host shared-memory channels without
sacrificing zero-copy performance. By treating the hypervisor as a trusted CA
and implementing a hypervisor-mediated mutual-authentication handshake, our
design prevents spoofing and impersonation attacks, while dynamic kernel-module
hooks enforce fine-grained channel access control. Microbenchmarks show that
the one-time handshake incurs less than 100 µs latency and that subsequent data
transfers achieve near-native IVSHMEM throughput and $\le{5\%}$
round-trip latency overhead. Thus, Secure IVSHMEM is well-suited for safety
critical, high performance environments such as automotive systems, where
untrusted services share memory in the same host.

\bibliographystyle{ACM-Reference-Format}
\bibliography{references}

\end{document}